%% file: Eds2017-AKK.tex
\documentclass[12pt]{article}
\usepackage{graphicx}

\def\uff{1 - Instituto de F\'isica - Universidade Federal Fluminense, Niter\'oi 24210-346, RJ, Brazil }

\def\ufrj{2 - Instituto de F\'isica - Universidade Federal do Rio de Janeiro, Rio de Janeiro 21945-970, RJ, Brazil }

\def\Title#1{\begin{center} {\Large #1 } \end{center}}
\def\Author#1{\begin{center}{ \sc #1} \end{center}}
\def\Address#1{\begin{center}{ \it #1} \end{center}}

\newenvironment{Abstract}{\begin{quotation}  }{\end{quotation}}
\newenvironment{Presented}{\begin{quotation} \begin{center} 
             Presented at\end{center}\bigskip 
      \begin{center}\begin{large}}{\end{large}\end{center} \end{quotation}}
\def\Acknowledgements{\bigskip  \bigskip \begin{center} \begin{large}
             \bf ACKNOWLEDGEMENTS \end{large}\end{center}}

\input econfmacros.tex

\begin{document}
\begin{titlepage}

\vfill
\Title{Description of pp forward elastic scattering at 7 and 8 TeV}
\vfill
\Author{A. K. Kohara$^{1}$, E. Ferreira$^{2}$, T. Kodama$^{1, 2}$ and M. Rangel$^{2}$}

\Address{\uff}
\Address{\ufrj}

\vfill
\begin{Abstract}
We analyse the recent LHC data at 7 and 8 TeV for pp elastic scattering with special attention for the structure of the real part, which  is shown to be crucial to describe the differential cross section in the forward region. We determine accurately the position of the zero of  the real amplitude, which corresponds to the zero of a theorem by A. Martin. 
\end{Abstract}
\vfill
\begin{Presented}
EDS Blois 2017, Prague, \\ Czech Republic, June 26-30, 2017
\end{Presented}
\vfill
\end{titlepage}
\def\thefootnote{\fnsymbol{footnote}}
\setcounter{footnote}{0}
%



The elastic amplitude $T(s,t)$ is a function of only two kinematical variables, controlled by principles of analyticity and unitarity, but no fundamental solution is known for its form, and representations of the differential cross section are given in terms of models, designed and applied for restricted ranges of $s$ and $t$. It is expected that at high energies the $s$ dependence becomes relatively simple, but the enormous gaps and uncertainties in the data from CERN/ISR, Fermilab and CERN/LHC do not help in tracing the $s$ dependence with reliability. On the other hand, for a given $s$, the angular dependence has not been measured with uniformity in the full $t$-range, and the necessary disentanglement of the real and imaginary parts of the amplitude is a hard task, with unavoidable indetermination \cite{LHC7TeV,LHC8TeV}. The forward $t$ range has been measured more often. Recently Totem and Atlas  groups at LHC measured $d\sigma/dt$ in forward $t$ ranges at $\sqrt{s}$ = 7 and 8 TeV \cite{T7, A7, T8, A8}. These data (Table \ref{datasets}) offer an opportunity to study in details several aspects of the very forward region, such as the magnitudes of the real and imaginary amplitudes, the position of the zero of the real part and the first derivatives of the amplitudes with respect to the variable $t$ (slopes). 
   The Coulomb-nuclear interference  depends on the proton electromagnetic structure, and the relative phase requires specific assumptions for the forms of the nuclear  amplitudes as described in details in our recent work \cite{us}. 
   

In the present work we propose independent parametrizations for the real and imaginary nuclear parts, writing
\begin{equation}
    T_R^N(t)= [1/(4 \sqrt{\pi} \left(\hbar c\right)^2)]~ \sigma (\rho -\mu_R t) ~ e^{B_R t/2}  ~,
\label{real_TR}
\end{equation}
and 
   \begin{equation}
T_I^N (t)= [1/(4 \sqrt{\pi} \left(\hbar c\right)^2)] ~ \sigma ( 1 -\mu_I t )~ e^{B_I t/2} ~ .
\label{imag_TI}
\end{equation}
The parameter $\sigma$ is the total cross section, $\rho$ is the ratio of the real and imaginary parts at $|t|=0$, $B_R$ and $B_I$ are the local slopes of the amplitudes and the parameters $\mu_R$ and $\mu_I$ account for the existence of zeros in the amplitudes. The zero in the real part is crucial to explain the $t$ dependence of $d\sigma/dt$ for small $|t|$.


The good quality of Totem data at 8 TeV in the forward region confirms that the differential cross section cannot be described by a pure exponential form like $d\sigma/dt= A\exp(B t)$. The non-exponential behaviour is obvious beforehand, since the  total cross sections is a sum of squared real and imaginary amplitudes with different slopes. The zero of the real part is  given  by  $t_R=\rho/\mu_R$. We understand that this quantity is the zero predicted in the theorem by A. Martin \cite{Martin}. \\

{\bf Data analysis  at $\sqrt{s}=$ 7 and 8 TeV}\\

The analysed datasets and  their $t$ ranges   are listed   in Table \ref{datasets}, 
where  T7, T8, A7, A8  specify  Totem (T) and Atlas (A) 
Collaborations and center-of-mass energies 7 and 8 TeV. 
In the measured ranges the Coulomb effects play 
important role and the relative 
Coulomb phase is properly taken into  account \cite{us}.

In order to identify  values for parameters valid for all measurements,  we study four different conditions  in the fits: 
   I) all six  parameters  are  free  ;  
 II) fixing $\rho$ at 0.14, as suggested by dispersion relations;  
 III) fixing $\mu_I$  from  the expected positions of   imaginary zero \cite{LHC7TeV, LHC8TeV} and dip in $d\sigma/dt$; 
    IV) fixing simultaneously $\rho$ and $\mu_I$ at the above values. A complete table with the  results can be found in ref. \cite{us}, and values obtained with Condition IV) are shown in Table \ref{Table:FINAL}.     Fixing both $\rho$ and $\mu_I$  at their {\it expected}  values  we obtain good modelling  for all measurements, except for the total cross sections, that separate Atlas from Totem.  
    
\begin{figure}
  \includegraphics[width=7.5cm]{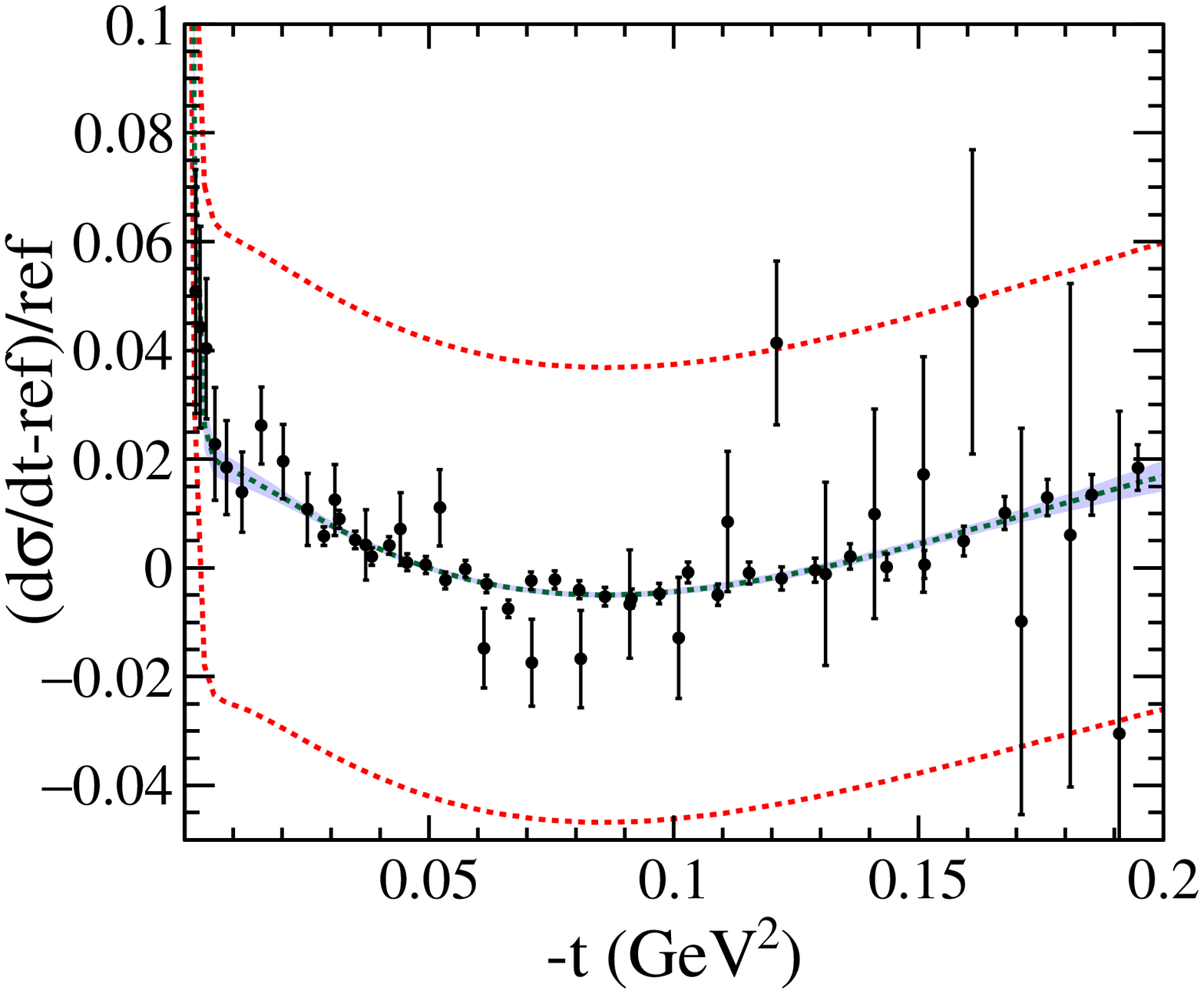} 
  \includegraphics[width=7.8cm]{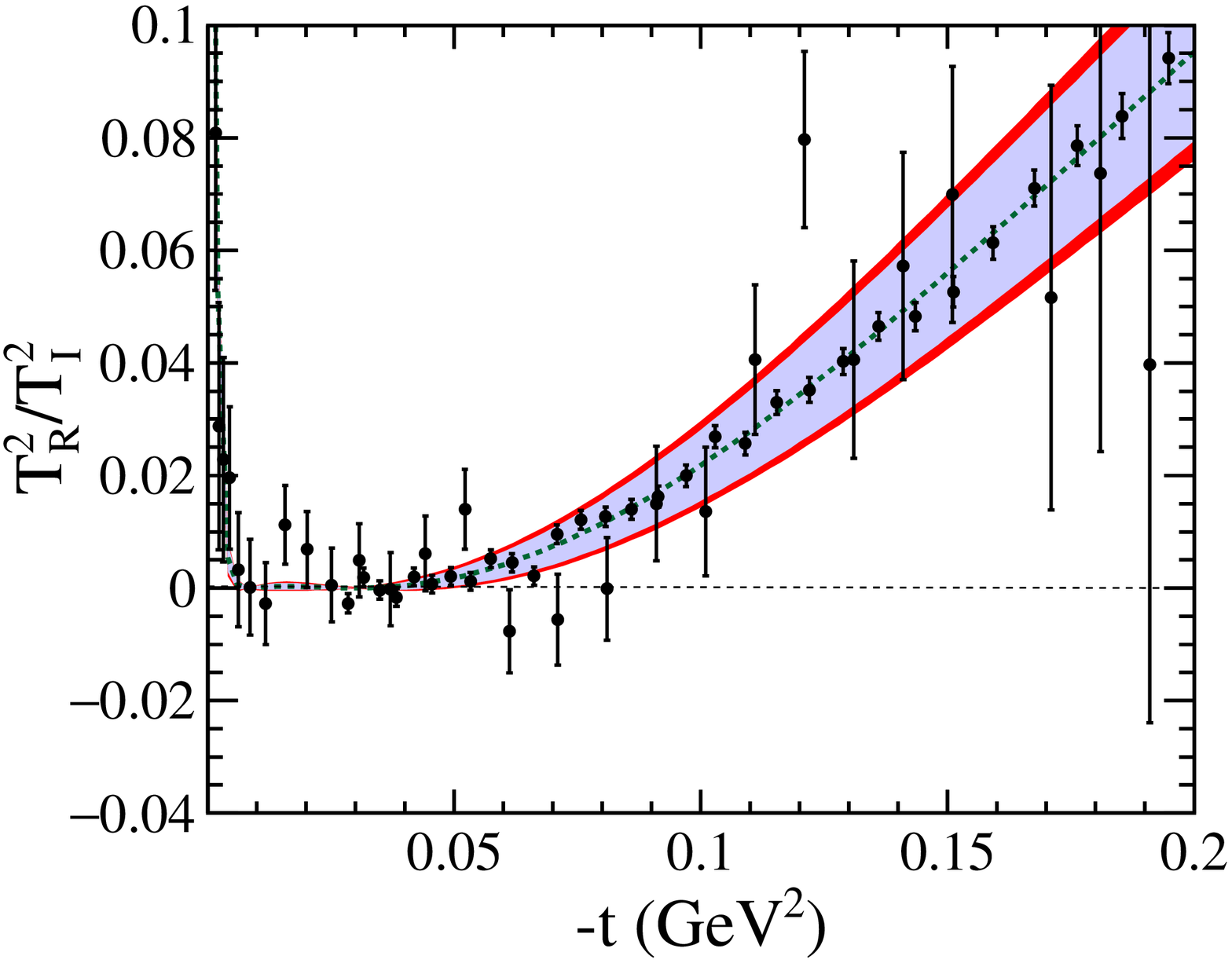}  
\caption{The left  plot  shows the non-exponential behaviour of the differential 
cross section for T8. The figure is obtained subtracting from the best fit  of the 
differential cross section a reference function which is $d\sigma/dt$ written 
 with a pure exponential form $ref = A\exp(B t)$  and dividing the subtraction by this reference function. 
The dashed lines show the normalization error band
in $d\sigma/dt$, that is quite large. The  plot in the RHS shows the ratio $T_R^2/T_I^2$ 
which exhibits information of a non-exponential behaviour with advantages compared with 
the first plot, since $\sigma$ is cancelled, and with it most of the normalization
systematic error.}
\label{displacement} 
\end{figure}

      The regularity on the values of $\mu_R$ is remarkable. 
 The position of the zero of the real part $t_R$ determined by $\rho/\mu_R$ is associated with the predicted zero of  A. Martin \cite{Martin},  is stable in all measurements with  $-t_R\simeq 0.037$ GeV$^{2}$ within the statistical errors. The position of the zero, together with the magnitude 
of $B_R$ determines the structure of the amplitudes shown in  Fig. \ref{displacement}. 

\begin{table*}
\begin{center}
 \vspace{0.5cm}
  \footnotesize
\begin{tabular}{ c c c c c c c c }
\hline
\hline
      &             &               &   &  &  &      &      \\
\hline 
  $\sqrt{s}$  & dataset  & $\Delta {|t|}$  range &  N         & Ref. &  $\sigma$ & $B_I$& $ \rho$        \\
     (GeV)    &               & (GeV$^{2}$)          &  points    &      &   (mb)    & ( GeV$^{-2})$ &     \\
\hline 
  7           &   T7       &  0.005149-0.3709      & 87         & 1    & 98.6$\pm$2.2      &19.9$\pm$0.3     & 0.14 (fix)$^{\rm a}$  \\
\hline
   7          &   A7       &  0.0062-0.3636        & 40        & 2     & $ 95.35\pm  0.38 $  & $19.73\pm 0.14 $ & 0.14 (fix) $^{\rm b}$       \\
 \hline
   8           &   T8       & 0.000741-0.19478      & 60       & 3     &  $103.0 \pm 2.3   $  &  19.56 $\pm$ 0.13     & (0.12 $\pm$ 0.03) $^{\rm c}$     \\
\hline
   8          &   A8       &   0.0105-0.3635       & 39        & 4     & $96.07\pm 0.18 $    &  $19.74\pm 0.05$     &  0.1362  (fix)$^{\rm d}$    \\
 \hline
 \end{tabular}
 \caption{ Values of parameters at $\sqrt{s}=$7 and 8 TeV determined by Totem and Atlas Collaborations at 
LHC \cite{T7,A7,T8,A8}. Values for $\rho^{\rm [a]}$ and $\rho^{\rm [b]}$   are taken from COMPETE Collaboration \cite{COMPETE};  $\rho^{\rm [c]}$  obtained by the authors with 
a forward SET-I and kept fixed in a complete  SET-II; $\rho^{\rm [d]}$ is taken from \cite{PDG}.}  
\label{datasets}
\end{center}
\end{table*}

\begin{table*}
\begin{center}
 \vspace{0.5cm}
 \footnotesize
\begin{tabular}{c c   c c c c c c c}
\hline
\hline
           &     &  &   &          &  &   &         \\
\multicolumn{8}{c}{Fixed Quantities :   $\rho = 0.14 $ , $\mu_I = - 2.16~ {\rm GeV}^{-2}$ (8 TeV) \cite{LHC8TeV}, $\mu_I = -2.14~ {\rm GeV}^{-2}$ (7 TeV) \cite{LHC7TeV} }  \\
\hline 
\hline 
&   N &$\sigma$       &$ B_{\rm I}  $      &$ B_{\rm R}  $      & $\mu_{\rm R}  $      & $-t_R$  &  $\chi^2/$ndf\\
  &         & (mb)                 &$({\rm GeV}^{-2})$ &$({\rm GeV}^{-2})$ & $ ({\rm GeV}^{-2})$ & $({\rm GeV}^{2})$ &         \\
      \hline
       \hline
T8&   60   &102.40$\pm$0.15& 15.27$\pm$0.39      &21.15$\pm$0.39      & -3.69$\pm$0.15      &   0.038$\pm$0.002 & 69.2/56  \\
  \hline
A8&  39   &96.82$\pm$0.11 & 15.26$\pm$0.06      &21.65$\pm$0.24      & -3.69$\pm$0.12      &  0.038$\pm$0.001  & 29.97/35  \\
 \hline
T7&   87    &99.80$\pm$0.21 & 15.71$\pm$0.14      &24.26$\pm$0.47      & -4.24$\pm$0.31  & 0.033$\pm$0.002 & 95.08/83  \\
\hline
T7& 87+17  &99.44$\pm$0.14 & 15.44$\pm$0.07      &22.62$\pm$0.19      & -3.49$\pm$0.13     & 0.040$\pm$0.002  & 203.5/100  \\
\hline
A7& 40   &95.75$\pm$0.16 & 15.23$\pm$0.11      &21.86$\pm$0.44      & -3.99$\pm$0.22      & 0.035$\pm$0.002  & 27.33/36  \\
\hline
\hline
\end{tabular}
 \caption{Proposed values of parameters for the  four datasets. The T7 data are also shown with inclusion of points at 
higher $|t|$ ($0.005149 < |t| < 2.443$ GeV$^2$)  that are important for confirmation of the value of $\mu_I$  \cite{us}. }
\label{Table:FINAL}
\end{center}
\end{table*}



The zero of the imaginary  part  anticipates the dip in the differential cross section that occurs beyond the range of the available data under study.

Our analysis indicates  that the real amplitude plays crucial role in the description of the  differential cross section in the forward region.
Interference with the Coulomb interaction is properly accounted for, 
and use is made of  information from  external sources, such as dispersion relations and predictions for the imaginary zero obtained in studies of full-$t$ behaviour of the differential cross section \cite{LHC7TeV,LHC8TeV}. 




\Acknowledgements 
AKK wish to thanks EDSBlois 2017 organizers for this stimulating Conference. This work is a part of the Brazilian project INCT-FNA Proc. No. 464898/2014-5. The authors wish to thank the Brazilian agencies CNPq, CAPES for financial support.

\end{document}

%% file: econfmacros.tex
\def\beq{\begin{equation}}
\def\eeq#1{\label{#1}\end{equation}}
\def\eeqn{\end{equation}}

\def\beqa{\begin{eqnarray}}
\def\eeqa#1{\label{#1}\end{eqnarray}}
\def\eeqan{\end{eqnarray}}

\let\bar=\overbar

\def\Dslash{\not{\hbox{\kern-4pt $D$}}}
\def\dslash{\not{\hbox{\kern-2pt $\del$}}}

\def\msb{{\bar{\ssstyle M \kern -1pt S}}}

